\documentclass{osa-article}

\journal{oe}

\articletype{Research Article}

\usepackage{mathtools}

\usepackage[toc,page]{appendix}
\usepackage{float}
\usepackage{amsmath}
\usepackage{longtable}
\usepackage{listings}
\usepackage{dirtytalk}
\usepackage{algorithm}
\usepackage{algpseudocode}
\usepackage{multirow}
\usepackage{fancyhdr}
\usepackage{lastpage}
\usepackage{graphicx} 
\graphicspath{ {figures/} }
 
\begin{document}

\title{Implementation of Optical Deep Neural Networks using the Fabry-P\'erot Interferometer}

\author{Benjamin David Steel\authormark{1,*}}

\address{\authormark{1}Computer Science and Electronics Student, University of Bristol, Bristol, UK}

\email{\authormark{*}bs15324@my.bristol.ac.uk} 



\begin{abstract}
Future developments in deep learning applications requiring large datasets will be limited by power and speed limitations of silicon based Von-Neumann computing architectures. Optical architectures provide a low power and high speed hardware alternative. Recent publications have suggested promising implementations of optical neural networks (ONNs), showing huge orders of  magnitude efficiency and speed gains over current state of the art hardware alternatives. In this work, the transmission of the Fabry-Perot Interferometer (FPI) is proposed as a low power, low footprint activation function unit. Numerical simulations of optical CNNs using the FPI based activation functions show accuracies  of 98\%  on the MNIST dataset. An investigation of possible physical implementation of the network shows that an ONN based on current tunable FPIs could be slowed by actuation delays, but rapidly developing optical hardware fabrication techniques could make an integrated approach using the proposed FPI setups a powerful solution for previously inaccessible deep learning applications.
\end{abstract}


\section{Introduction}
\pagenumbering{arabic}

Deep learning techniques have resulted in significant performance gains in pattern recognition problems in the last decade, in fields from image and speech recognition to language translation \cite{DeepLearning, SilverHuangEtAl16nature, mnih2015humanlevel, Krizhevsky:2012:ICD:2999134.2999257}. 

However, deep neural networks modelled on a traditional general purpose Von-Neumann computer result in slow and power hungry performance due to the computation bottleneck of data movement \cite{Esser11441}. Recent development in hardware architectures tailored towards deep learning applications has allowed performance gains, using hardware such as GPUs, application-specific integrated circuits (ASICs) and field-programmable gate arrays (FPGAs) \cite{Esser11441, 58356, 10.3389/fnins.2011.00108, misra2010artificial, chen2017eyeriss, graves2016hybrid, AccelNNAnalogue}.

An alternative artificial neural network (ANN) accelerator is a specialised optical hardware implementation: an optical neural network (ONN). The ANN computing paradigm is well suited to a fully optical system, with simulations of ONNs suggesting 10 orders of magnitude efficiency improvements and 3 orders of magnitude speed improvements over NVIDIA GPUs \cite{miscuglio2018all}.

ANNs rely heavily on fixed linear transformations, which can be performed at the speed of light in optics and detected at rates of up to 100GHz in photonic networks \cite{vivien2012zero}, with minimal power consumption \cite{yang2013multiplier}. ONNs also allow greater neuron fan-in/fan-out due to lack of electromagnetic compatibility (EMC) concerns, allowing superior connectivity and scalability compared to silicon based ANNs \cite{40669}.

Hybrid ONNs implemented in free-space and integrated systems have been shown to be successful, but don't fully provide promised optical power and speed advantages due to the time consuming process of non-local digital computation of activation functions \cite{krishnamoorthy1992scalable, DBLP:journals/corr/abs-1901-03661, HybridCNNs, chen2016asp, Hughes:18, hamerly2018large}. Fully optical ANNs have also been demonstrated, but for now require offline training or are not successfully implemented in hardware \cite{DeepLearningNanophotonicCircuits,miscuglio2018all, bueno2018reinforcement}. Industry is also making steps towards an ONN, in companies such as Fathom Computing, Lightelligence and Optalysis \cite{FathomComputing, Lightelligence, Optalysys}.

The use of digital computation to implement a standard activation function such as \textit{ReLU} in many of the above networks locks the ONN to the relatively slow digital processor GHz-scale clockrates, and the limited number of input and output channels limits the scalability of the design. These functions are difficult to implement in optics while still maintaining low power consumption and a small physical footprint \cite{2019arXiv190304579W}. 

This work proposes a system of local intensity measurements with no data movement, a propagated optical signal, combined with an optical non-linearity to achieve a high speed optical non-linear layer. This can be achieved using a Micro-Electro-Mechanical System (MEMS) based tunable Fabry-Perot Interferometer (FPI), which features a non-linear frequency transmission, has low power requirements, and has very high sensitivity allowing low power optical sources \cite{wildfeuer2009resolution}. The field transmission of the FPI, taken from ref. \cite{Gorodetksy:10}, is as shown in equation \eqref{eq:FPIFieldTransmission}. 

\begin{equation}
    T(\omega) = \frac{\kappa}{-j(\omega - \omega_c) + \kappa}
    \label{eq:FPIFieldTransmission}
\end{equation}

Where $\omega$ is the frequency of the input electromagnetic wave to the FPI. $\omega_c$ is a tunable electronic input that can modulate the centre frequency of the device, thereby allowing it be used as a tunable weight, and $\kappa$ relates to the full width at half maximum linewidth $\Delta v_c$, a fixed variable for an FPI array \cite{hecht2013optics}. Combined with a complementary metal-oxide semiconductor (CMOS) image sensor, the FPI transmission can be used as a function of optical amplitude instead of optical frequency. 

Two suggested ONN architectures will be modelled in this work. First an incoherent simulation will be numerically simulated, where the propagating optical signals consist of fields with a non-constant phase difference. Second a coherent simulation will be tested, where the optical signals consist of fields with a constant phase difference. Complex values will be used in this network, with complex multiplications, and therefore correspondingly the complex valued field transmission will be used as opposed to intensity transmission.

The main application of this research is in the field of computer vision, due in part to the classification of multi-dimensional images being a task inherently suited to a free-space optical system due to its naturally high degree of 2D connectivity \cite{1057566}. As such, a CNN architecture will be used in this work, a class of ANN used in tasks such as image recognition, segmentation and generation \cite{MAL-006, Long_2015_CVPR, goodfellow2014generative, Krizhevsky:2012:ICD:2999134.2999257}.

There are however some key departures from conventional CNN design that the optical CNN will be required to take. Traditional CNNs typically use a max function in a \textit{pooling layer} to improve generalisation, which is difficult to implement in optical hardware \cite{2019arXiv190304579W}. However, it has been shown that this function is not strictly necessary, provided the stride length/kernel size of the network can be increased \cite{2014arXiv1412.6806S}.

\section{Activation Function}

To take full advantage of the optical non-linearity, the optical signals in the network will be propagating through the FPI. This will ensure fewer required optical sources and therefore a faster and more compact ONN. This results in an effective transfer function of $xT(x)$ where $x$ is the intensity or field passing through the FPI.

This activation is simply a function of the intensity of the signal, and therefore will be used in the incoherent model of the ONN. Note that the input $x$ can only assume positive values due to it being the intensity of light. If it is assumed that the input operates in a positive range above zero, the function  tends back to the shape of the Lorentzian for increasing intensities. Therefore henceforth the Lorentzian function will be used in incoherent models to allow less numerical complexity and faster simulation.

In the coherent model, as before propagated light will result in a $xT(x)$ activation function, where $T(x)$ is the field transmission in equation \eqref{eq:FPIFieldTransmission}, and $x$ is a complex valued light field. CMOS detection will be modelled with a $\Re(xx^*)$ term.

Typically activation functions are required to be monotonic, which ensures a convex error surface and therefore increased ease of convergence. However, recently suggested activation functions like \textit{Swish}, are non-monotonic and have been shown to outperform \textit{ReLU} on many datasets \cite{DBLP:journals/corr/abs-1710-05941}.

Approximating the identity function near the origin allows a simple initialisation of 0 for the bias weight. The field transmission fulfills this requirement, but the intensity transmission does not. Therefore care will be need to be taken when initialising the bias weight for the incoherent model \cite{sussillo2014random}.

Non-saturating activation functions such as \textit{ReLU}, have been shown to converge faster \cite{Krizhevsky:2012:ICD:2999134.2999257}. Saturating activations such as the logistic, tanh and the Lorentzian activations presented here have a diminishing gradient as the input tends to $\pm \infty$, resulting in neurons that change increasingly slowly.

In this implementation the electronic bias parameter $x_0$ that actuates the FPI will also be used as the bias parameter typical in ANNs. This has the advantage of requiring less optical hardware than another optical signal based weight. Trainable parameters of the activation function have a strong precedent in ANNs \cite{DBLP:journals/corr/abs-1710-05941, DBLP:journals/corr/abs-1710-05941}. 

For the intensity activation function that is simply the Lorentzian function, changing the bias parameter $x_0$ equates to a lateral shift of the function. Increasing the bias parameter $x_0$ of the complex valued field activation function increasing the order of the function, therefore increasing the number of turning points.

Increasing $\kappa$ for both functions results in a decrease in gradient and range, and therefore an optimal value will likely be linked to the learning rate of the network.

The initialisation used in this work for linear transform weights is the \textit{normalized initialisation}, reduces the risk of saturation of the activation function, an important consideration considering the activation function presented here is saturating \cite{glorot2010understanding}. The bias is normally initialised as 0, but due to the activation function presented here having a low gradient around the origin, the bias is initialised with a uniform distribution in a constant range to try and increase the convergence rate \cite{glorot2010understanding}. The range that works best for this initialisation will have to be experimentally discovered.

\section{Incoherent Model}

In this work several datasets will be used to fully test the performance of the proposed ONN architecture. CNNs are incredibly flexible, and have many possible configurations in terms of size and order of layers. These advantages and disadvantages of different model layouts for the unique activation function being introduced will be explored with several image datasets such as the Semeion digit dataset \cite{SemeionDataset}, and various shape datasets \cite{FourShapesDataset, ThreeShapesDataset}.

\textit{Python} with \textit{NumPy} was used to develop initial models \cite{Python, NumPy}. Then later when very high performance was needed for large datasets such as MNIST, a \textit{TensorFlow} implementation was built \cite{TensorFlow}.

\subsection{Implementation}

The stochastic gradient descent optimizer would likely by the easiest to implement in hardware. However for the sake of numerically modelling the ONN on a computer, an optimizer that allows faster convergence would allow more results to be obtained. For this reason a mini batch Adam optimizer was used \cite{kingma2014adam}. It would likely be difficult to synthesize in hardware for initial ONN implementations, but the promised speed up of optics should allow a simple gradient descent to be fast enough to beat a digital electronics based Adam optimizer anyway.

The commonly used mean squared error (MSE) loss function would likely be easiest to implement in optics if an all optical in-situ trainable solution was required. However to ensure faster training for the numerical simulations, the \textit{categorical cross entropy} loss function was combined with the \textit{softmax} function, a combination better suited to classification problems \cite{doi:10.1162/089976604773135104}.

\subsection{Testing}

\subsubsection{Impact of Hyperparameters}

Two datasets were used for initial experimentation, the Iris flower classification dataset, consisting of 3 classes with 50 examples each \cite{IrisDataset}, and the larger and more difficult UCI heart disease dataset \cite{HeartDiseaseDataset}, a binary classification problem consisting of 300 patients with 13 attributes each.

To explore how the value of $\kappa$ effects the accuracy and convergence speed of the network, $\kappa$ was varied by layer for a range of values, for a learning rate of 0.01.

A model with two hidden layers was used, and for this experiment was run with combinations of layer sizes of 16 and 64. This was done to try to understand if there was an ideal $\kappa$ value for a certain ratio of layer size, due to the relationship between $\kappa$ and the range of the Lorentzian function.

Experimentation shows that the network seems to be insensitive to the first layer $\kappa$ value, but requires a relatively low second layer $\kappa$. It can also be seen that larger values of $\kappa$ greatly reduce the rate of convergence for the model. There seems to be no significant relationship between $\kappa$ value and layer size ratio from this data.

\begin{figure}
    \centering
    \includegraphics[width=0.8\textwidth]{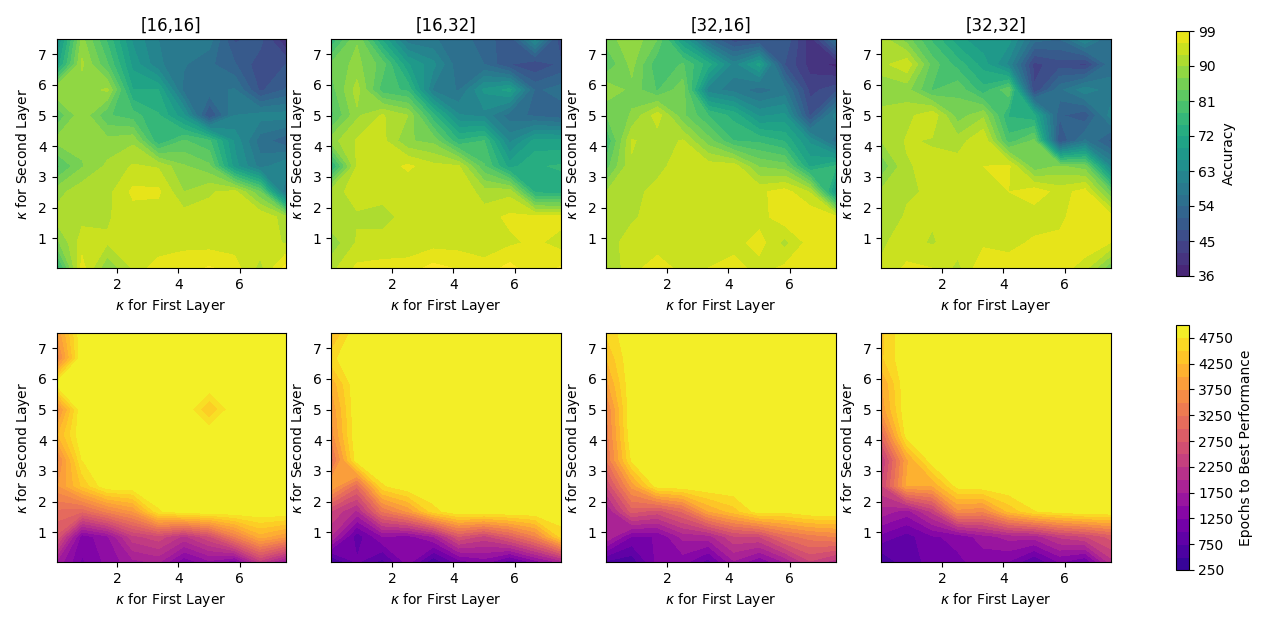}
    \caption{An exploration of how altered $\kappa$ by layer effects the accuracy and convergence speed of the incoherent model on the heart disease dataset. The accuracy of the network appears to be insensitive to a range of $\kappa$ for subsequent layers of a range of size ratios, other than requiring a relatively low summed value. It can be seen that larger values of $\kappa$ quickly decrease network convergence speed.}
    \label{fig:IncoherentHeartLayerGamma}
\end{figure}

The experiment is repeated using the heart disease dataset. Fig. \ref{fig:IncoherentHeartLayerGamma} seems to suggest that the accuracy is insensitive to differing values of $\kappa$ at each layer, and a low value relative to the learning rate is required for a reasonable convergence rate. This is promising, as a hyperparameter needing to be tuned by layer would have an increased implementation and design complexity, and decreased generalisation ability.

To confirm the prior assumption that these trends are relative to the learning rate, the same trial of varying $\kappa$ by layer and plotting accuracy and convergence speeds was done for a range of learning rates. The data backs up the hypothesis that the $\kappa$ value relationships persist proportionally to the learning rate. This is promising in that for a fixed $\kappa$ defined by fabrication and implementation specifics, the learning rate can be simply be tuned to reach peak performance. A general rule of thumb seems to be a the learning rate needs to be 100 to 200 times less than $\kappa$ to achieve the high accuracy and convergence rates.

\begin{figure}[t]
    \centering
    \includegraphics[width=0.4\textwidth]{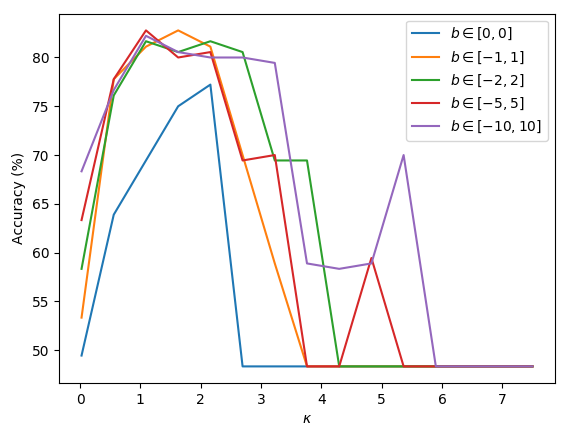}
    \caption{An exploration of how changing the range of the uniform distribution from which the initial $x_0$ are randomly chosen impacts the accuracy of the network, as $\kappa$ is varied. The heart disease dataset was used.}
    \label{fig:IncoherentHeartBias}
\end{figure}

With the introduction of a new activation function it was important to understand how the initialisation of the bias weight could impact the network's capabilities, especially in the case as here where the gradient is low around the origin \cite{glorot2010understanding}. Figure \ref{fig:IncoherentHeartBias} shows how changing the range of the uniform distribution from which the initial $x_0$ bias values are randomly chosen impacts the accuracy of the network, as $\kappa$ is varied.

The results indicate that for the \textit{Lorentzian} function, a wider spread of initial bias values can result in a larger insensitivity to the $\kappa$ value used on each layer. Any initialisation variance over zero seems to produce a $\kappa$ range of 100 to 300 times the learning rate where high accuracies are achieved.

\subsubsection{Kernel Size}

In an ONN the normal constraints of using smaller kernels are removed due to a constant time convolution operation made possible by free-space optics instead of a higher order time complexity operation as required in digital processing. Therefore there is more freedom of choice when it comes to choosing a kernel size. To explore this, the three shapes dataset \cite{ThreeShapesDataset} was used. 

\begin{figure}[h!]
    \centering
    \includegraphics[width=0.7\textwidth]{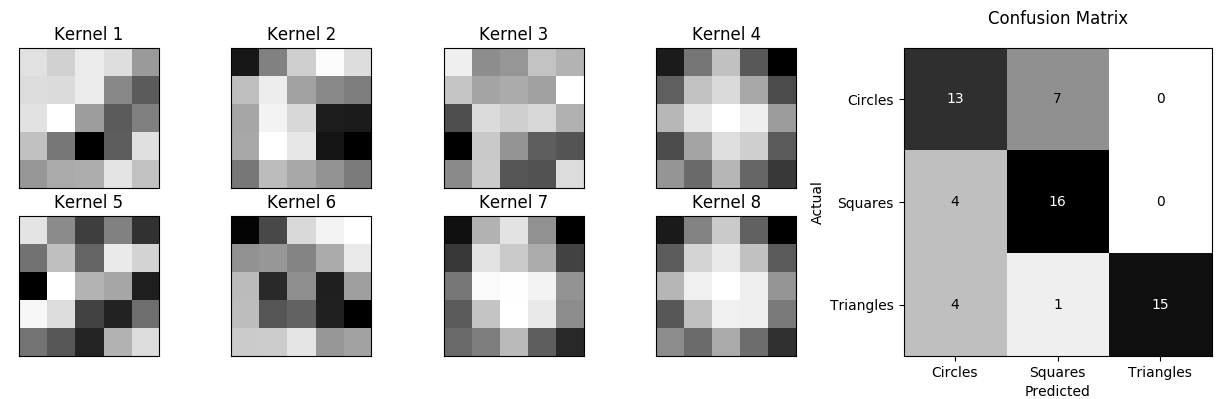}
    \caption{A visualisation of the first layer of convolutional kernels from a model trained on the three shapes dataset. The model architecture is 8 5$\times$5 convolutional filters for both the first and second layers, then a 128 neuron fully connected layer. This model achieved a 73.3\% accuracy on the test set. Note kernels 1-3 and 5-6 seem to have assumed diagonal edge detection kernels, while 4, 7 and 8 have assumed very similar kernel patterns.}
    \label{fig:IncoherentThreeShapesKernel5ConfMat}
\end{figure}

\begin{figure}[h!]
    \centering
    \includegraphics[width=0.7\textwidth]{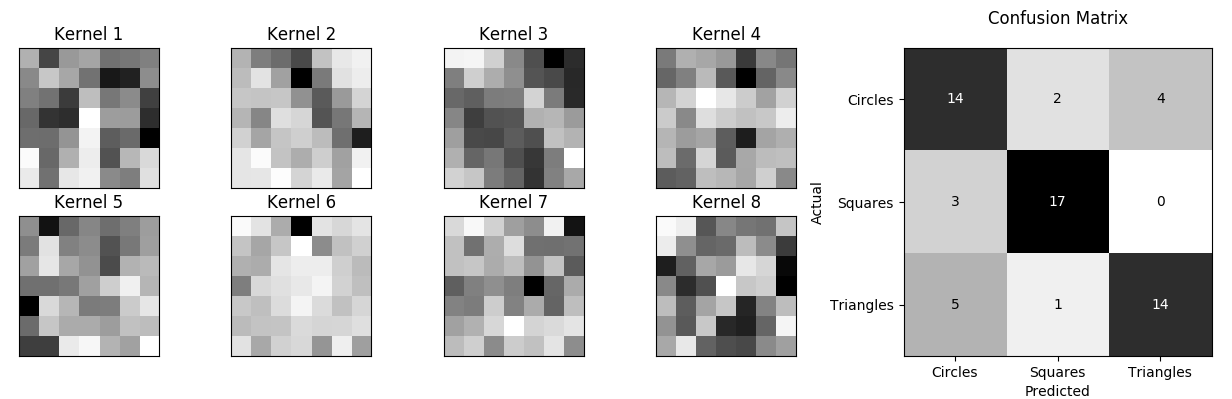}
    \caption{A similar visualisation as \ref{fig:IncoherentThreeShapesKernel5ConfMat}, but with a kernel size of 7$\times$7. This model achieved a 75\% accuracy on the test set. Note kernels 2-3 and 8 seem to have assumed diagonal edge detection kernels.}
    \label{fig:IncoherentThreeShapesKernel7ConfMat}
\end{figure}

Three different kernel sizes were tested: 3$\times$3, 5$\times$5 and 7$\times$7. For each of these sizes two models were trained. Fig. \ref{fig:IncoherentThreeShapesKernel5ConfMat} shows the first layer kernels and confusion matrix from the 5$\times$5 model. Fig. \ref{fig:IncoherentThreeShapesKernel7ConfMat} represents the 7$\times$7 model. 

The best performing network had kernels dimensions of half the size of the input images, a relatively large factor. However this could be due to the large nature of the features found in this dataset.

\subsubsection{MNIST Benchmarking}

The model was tested using the MNIST dataset. All CNN configurations used a single fully connected layer of 512. The model was tested for a variety of kernel sizes and pooling styles. Each configuration was tested three times, and the best error rate achieved was recorded as shown in table \ref{table:IncoherentMNIST}.

\begin{table}[h!]
\centering
\begin{tabular}{| c | c || c | c | c | c|} 
 \hline
 \multicolumn{2}{|c||}{\multirow{2}{*}{Error Rate (\%)}} & \multicolumn{4}{c|}{Kernel Size} \\
 \cline{3-6}
 \multicolumn{2}{|c||}{} & 3 & 5 & 7 & 9 \\
 \hline\hline
 \multirow{2}{*}{No. Pooling Layers} & 1 & 1.95 & 2.15 & 1.95 & 2.34 \\ 
 \cline{2-6}
 & 2 & 2.34 & 2.34 & 2.73 & 2.34 \\
 \hline
\end{tabular}
\caption{Best of 3 error rates for a incoherent numerical simulation of the FPI based ONN, implemented in \textit{TensorFlow} and trained on the MNIST dataset. The single pooling layer configuration consisted of two convolutional layers of stride 1 followed by a convolution of kernel size 4 and stride 4. The 2 pooling layer configuration had the two convolution layers sandwiched between pooling layers of stride 2. The single pooling layer can be seen to produce higher accuracies, while kernel size has little affect.}
\label{table:IncoherentMNIST}
\end{table}

\section{Coherent Modelling}

The next step in this work is to trial the field transmission function of the FPI. Optics is unique in that the signal passing through the network is inherently complex. Many optical components can implement an arbitrary phase shift (such as metasurfaces and MZIs, already used in ONN implementations), which means a complex matrix multiplication should be possible. 

Complex valued ANN have started to gain ground in the field of deep learning \cite{danihelka2016associative, 941159}, especially so in CNNs \cite{tygert2016mathematical, trabelsi2017deep}. There have also emerged strong potential uses in recurrent neural networks (RNNs) \cite{arjovsky2016unitary, jing2018gated,jose2017kronecker} and generative adversarial networks (GANs) \cite{mescheder2017numerics}.

\subsection{Testing}

The network showed extremely rapid convergence on many datasets, but poor generalization. This could potentially be due to the initial use of the networks of the same number of neurons as for the incoherent case, resulting in a doubling of parameters that tends to allow overfitting. For this reason larger datasets were used as a form of regularization, such as a dataset of four shapes consisting of 5000 examples \cite{FourShapesDataset}.

\subsubsection{Impact of Hyperparameters}

The coherent model seems to be insensitive to $\kappa$ being varied by layer for different layer size ratios. This is again good for an implementation, as a constant $\kappa$ providing the best performance allows a more general hardware implementation.

Due to the coherent transmission activation function approximating the identity at the origin, the initialisation of the bias weight $x_0$ should be less important \cite{sussillo2014random}. Experimentation shows that varying $\kappa$ and the initialisation variance seems to have little effect on the accuracy of the network, but that it is important to ensure the variance is limited to around 4 times $\kappa$ to ensure greater convergence rates.

Of note is that a high initialisation variance produces a significantly higher accuracy. This could be the result of the highly linear nature of the complex activation near the origin. A larger initialisation variance could allow the non-linearities of the activation function to be more easily exploited.

\subsubsection{Training Analysis}

To understand the impact each layer has on the output of the network, the distribution of activation function output values can be saved after each weight update, for each layer \cite{glorot2010understanding}. The experiments done here used the Semeion dataset.

\begin{figure}[h!]
    \centering
    \includegraphics[width=0.8\textwidth]{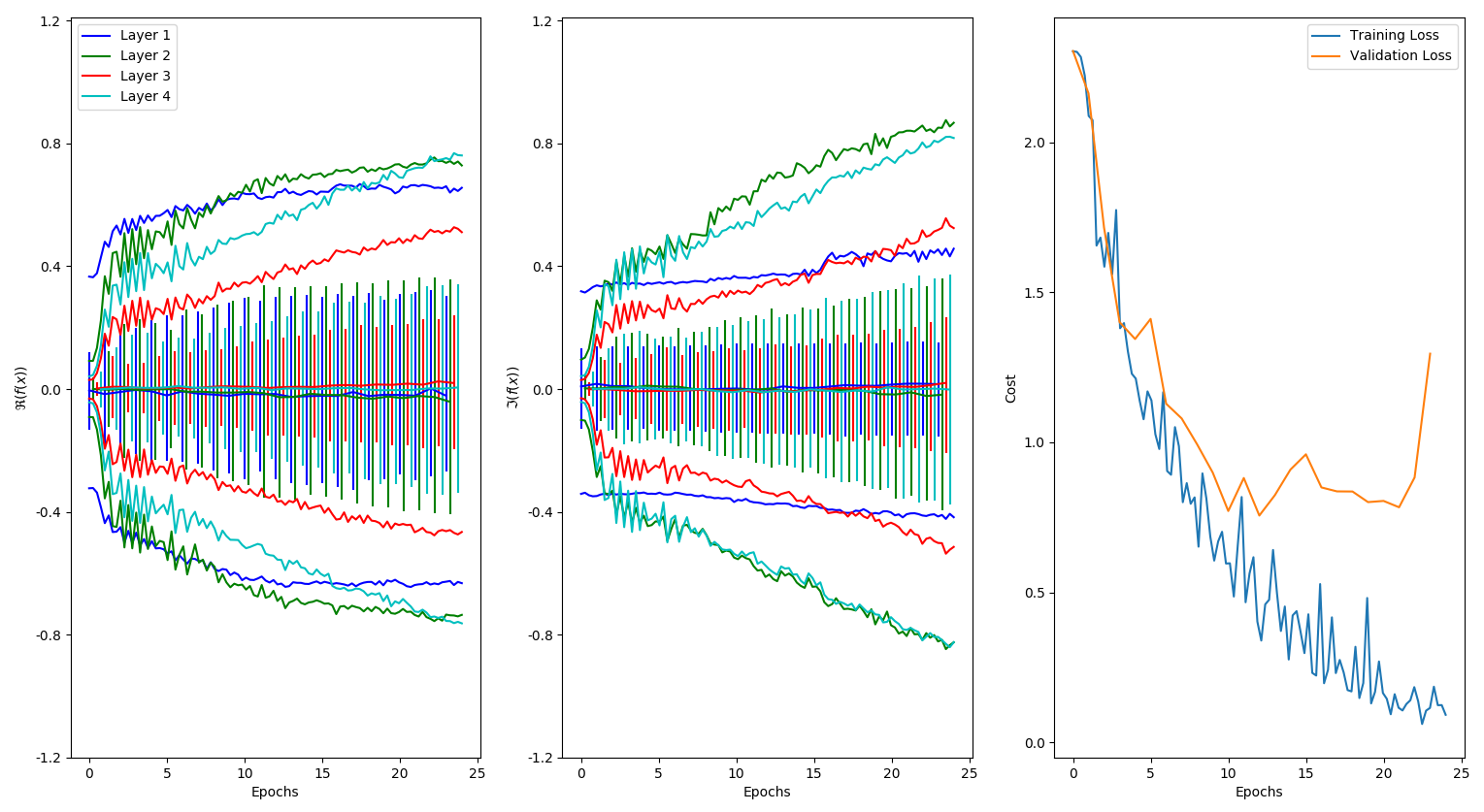}
    \caption{Plots on left and middle show distribution of activation function outputs after each weight update, with plot on right showing training and validation set loss for corresponding weight updates. Central line in distribution plots shows the median, with error bars showing the 1st and 3rd quartiles. Extreme lower and upper plots show 5\% and 95\% percentiles respectively. Data shown for model with $\kappa$ of 2 and initial bias spread of $[-0,0]$. }
    \label{fig:4Gamma0BiasSemeionByLayer}
\end{figure}

Figure \ref{fig:4Gamma0BiasSemeionByLayer} shows a visualisation of the training period of a network with the bias variable initialised to 0. The trained network achieved an accuracy on the test set of 79.3\% after 10 epochs.

\begin{figure}[h!]
    \centering
    \includegraphics[width=0.8\textwidth]{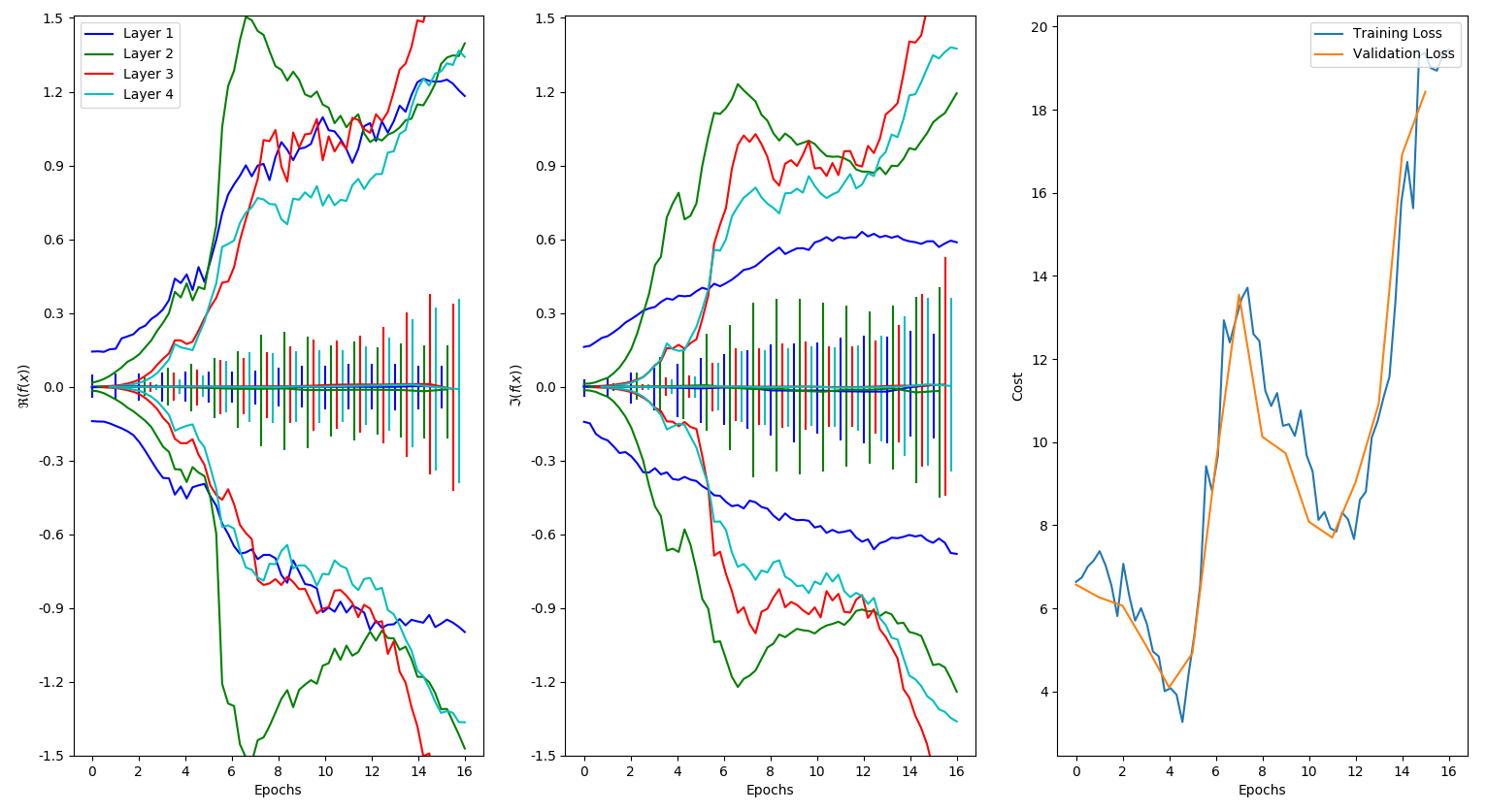}
    \caption{The subplots represent a visualisation of the training of the network in the same representation as figure \ref{fig:4Gamma4BiasSemeionByLayer}. Data shown for model with $\kappa$ of 2 and initial bias spread of $[-4,4]$.}
    \label{fig:4Gamma4BiasSemeionByLayer}
\end{figure}

Figure \ref{fig:4Gamma4BiasSemeionByLayer} again shows the same visualisation but with an initial bias variance of $[-4,4]$. The trained network only achieved a final accuracy on the test set of 39.0\% after 4 epochs. Extremely rapid growth of the activation output distribution range has resulted from the large initial bias variance. The unstable sinusodially varying range of activation outputs implies the reason for the instability is the non-monotonic nature of the activation function.

\subsubsection{Complex Kernels}

\begin{figure}[h!]
    \centering
    \includegraphics[width=0.7\textwidth]{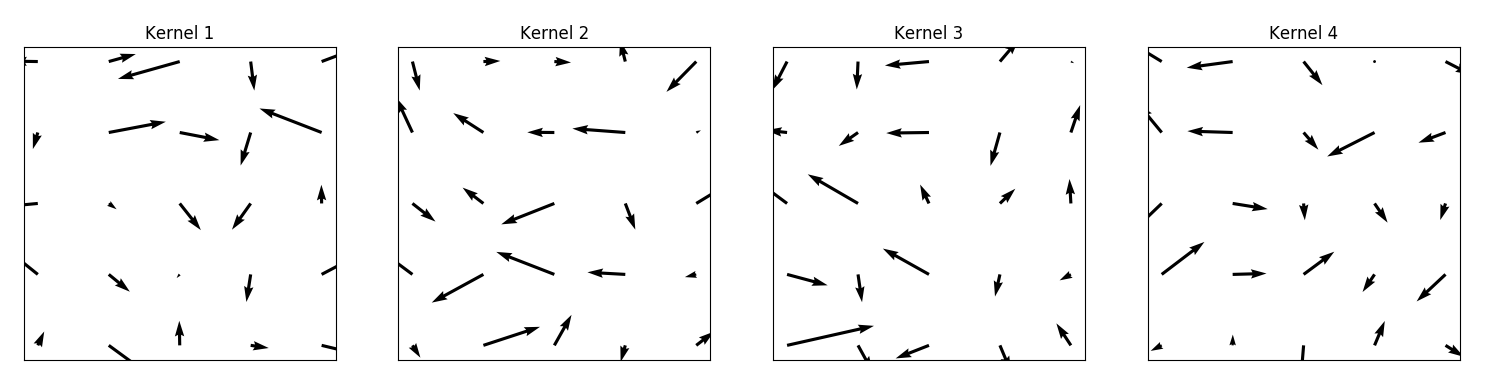}
    \caption{Kernels from a coherent model trained on a dataset of four shapes: triangles, circles, squares and stars. The first layer was comprised of 4 5$\times$5 complex valued kernels.}
    \label{fig:4ComplexKernelsFourShapes1}
\end{figure}

Fig. \ref{fig:4ComplexKernelsFourShapes1} shows a visualisation of the first layer of 4 5$\times$5 complex valued kernels, from a model trained on the four shapes dataset to an accuracy of 99.8\% \cite{FourShapesDataset}.

The kernels can be seen to have learned the features of the shapes they are trained on. Exploiting the complex nature of the fields in the coherent network has essentially given the model double the number of parameters for the same network size.

The same experiemnt was done on a first layer of complex kernels from a model, again trained on the Four Shapes dataset, but this time with 8 5$\times$5 kernels. The network achieved an accuracy of 99.9\%. It is notable that the two models achieved comparable accuracies despite the latter network being larger, which could be due to smaller networks being able to generalise better \cite{lecun1989generalization}.

\subsubsection{MNIST Benchmarking}

The coherent model was trained on the MNIST dataset using a TensorFlow implementation. A network of reduced size was used compared to the incoherent model, in order to keep a similar number of parameters, and therefore improve generalization.

The network consisted of 3 convolutional layers with the final layer acting as a pooling layer with higher stride, and a single fully connected layer. A $\kappa$ value of 8 was used, with a bias parameter initialisation variance of 0. The best performing network achieved an accuracy of 95.1\%.

The accuracy is lower than for the incoherent model, but this could be due to the tendency of complex valued ANNs to converge to a solution less often. Ref. \cite{guberman2016complex} only achieved high accuracies in MNIST after 20 iterations, with many networks failing to converge. Unfortunately due to computation time constraints not as many networks could be evaluated here.

\begin{figure}[h!]
    \centering
    \includegraphics[width=0.5\textwidth]{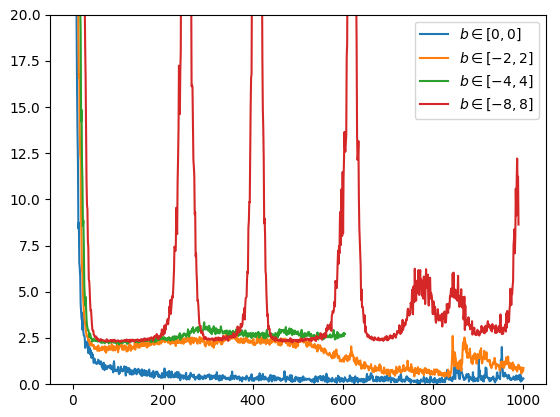}
    \caption{4 networks trained on the MNIST dataset. A network architecture of three convolutional layers with last layer acting as a pooling layer was used, the same architecture as described earlier in this section. A $\kappa$ value of 8 was used. The bias parameter initialisation variance is shown to be varied.}
    \label{fig:CohMNISTErrorCurves}
\end{figure}

Fig. \ref{fig:CohMNISTErrorCurves} shows validation loss over epochs for 4 networks trained on the MNIST database. A higher bias initialisation variance can be seen to create periodic instability, similar to that seen previously in the training analysis.

\section{Hardware Modelling}

To provide one example of a low level hardware modelling of the presented FPI activation function, the Python software packaged \textit{Neuroptica} \cite{neuroptica} was used. This is a photonic ONN simulator, allowing optical matrix multiplication numerical simulation down to the MZI and phase shifter level.

The library was extended by the author to support a trainable activation function, and once this was done the FPI based activation function was added. A peak accuracy of 85\% on the heart disease dataset was achieved. Experimentation here again shows that networks with a higher bias parameter initialisation variance have an unstable training stage. A higher $\kappa$ can also increase convergence speed.

The performance of the \textit{Neuroptica} ONN was not tested with an image dataset due to hardware limitations.

\section{Physical Implementation}

\subsection{FPI Integration}

\begin{figure}[h]
    \centering
    \includegraphics[width=0.5\textwidth]{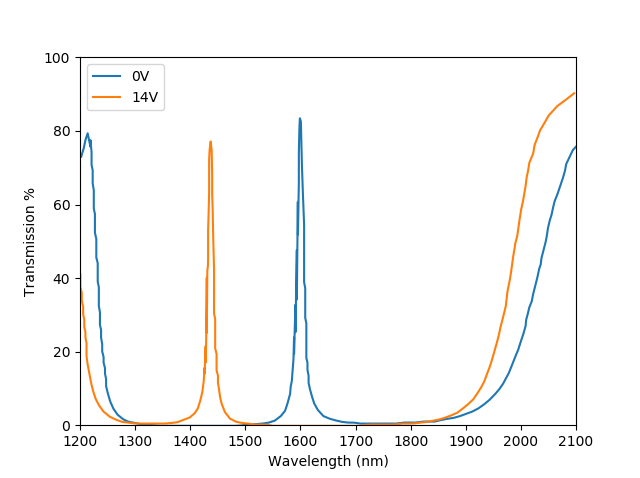}
    \caption{The intensity transmission of a MEMS actuated FPI at 2 voltage offsets is shown here. The frequency at which peak transmission is reached can be seen to be easily modulated. Data taken from ref. \cite{antila2010mems}.}
    \label{fig:VTTFPITransmission}
\end{figure}

Advances in Micro-Opto-Electro-Mechanical systems (MOEMS) FPI  technology and improvements in fabrication techniques have resulted in highly compact tunable microspectrometer arrays designed for visible and IR light, which have a low manufacturing cost and small physical footprint \cite{CORREIA2000191, MicrospectrometerArray, MEMS-FPItunable, blomberg2010electrically, RISSANEN2012130}. Huge growth has been seen in the last 10 years of patents using such devices \cite{antila2010mems}. The transmission of a MOEMS actuated FPI is shown in fig. \ref{fig:VTTFPITransmission}. 

The proposed implementation of the FPI to create an activation function that is dependent on intensity and not frequency involves the use of a CMOS light sensor detecting the optical signal passing through the FPI and modulating the centre frequency of the FPI. If the wavelength of the optical signal is kept constant this would result in an intensity transmission that matches the frequency transmission shown here. This design requires a CMOS sensor placed near enough to the FPI that they both receive an optical signal of the same intensity, phase and frequency.

Alternative architectures of a FPI based ANN designed with high connectivity in mind were developed, but ultimately abandoned, as the hurdle of inviting adoption of optical hardware by deep learning researchers will be more difficult if the designs that are offered are divergent from commonly accepted and successful mainstream architectures.

\subsection{Free-Space Approach}

Modern approaches using refractive optics have used \textit{4f correlators}. These systems use 2 lenses of equal focal length spaced 2 focal lengths apart, with input and output planes located at the front and back focal planes of the first and second lenses respectively. If a complex transmittance mask is used between the two lenses, an arbitrary matrix multiplication to be performed. It is incredibly fast and an in-situ retrainable solution, but is still somewhat bulky. It has however been used successfully to produce high accuracy results in hybrid optical CNNs \cite{DBLP:journals/corr/abs-1901-03661} and fully optical RNNs \cite{bueno2018reinforcement}.

Metasurfaces are a form of subwavelength diffractive optics technology that allow an arbitrary phase shift function to be implemented, compared to a simple quadratic phase shift function in refractive lenses. This allows linear transformations in an extremely compact form factor \cite{arbabi2015dielectric}. However once fabricated metasurfaces perform a fixed transformation, and therefore not particularly useful in the pursuit of an in-situ retrainable ONN.

Digital micromirror devices (DMDs) are a compact and high resolution MOEMS based array of microscopic mirrors that can be individually actuated \cite{MicromirrorArray}. They have been used in combination with $4f$ correlators to produce in-situ retrainable ONNs \cite{bueno2018reinforcement}.

\subsection{Performance}

\subsubsection{Speed}

Many of the operations required in an ANN are extremely computationally intensive. Linear transformation and convolution have a complexity of $O(n^3)$ and $O(n^2k^2)$ (where $n$ is the input dimension and $k$ is the filter dimension) respectively. This can be compared with a basic $4f$ correlator which can perform an arbitrary matrix multiplication or convolution with complexity $O(1)$ where computation time is simply the time taken for light to travel the length of the correlator. Metasurfaces could reduce this time further.

The propagation time of the FPI is the time it takes for a single round trip of light in the device for contructive interference to occur, which is $T_{FPI} = \frac{2l}{c}$ where $c$ is the speed of light, and $l$ is the geometrical length of the FPI. Therefore the total time for optical signals to propagate through the system is roughly $T_{total} = L(T_{MatMul} + T_{FPI})$ where $L$ is the number of layers in the network, assuming the correlators and activation functions could be stacked adjacently.

The maximum frequency the MOEMS based tunable FPI can be actuated at is 5.1kHz, from minimum to maximum air gap, corresponding to a step response delay of 31$\mu$s \cite{blomberg2010electrically}. This means the FPI actuation speed is by far the prevalent delay in the network, resulting in a worse case forward propagation time of $T_{total} = 31L\mu s$ where L is the number of layers in the network.

\subsubsection{Power}

The energy efficiency of ANN hardware implementations is typically measured in multiply accumulate (MAC) operations per Joule (MAC/J), which is proportional to the number of neuron operations per Joule. GPU based ANNs typically have an efficiency of $10^{6}$ MAC/J, electro-optical ONNs have shown efficiency of $10^{12}$ MAC/J, and ONNs using passive optical activation functions have shown $10^{16}$ MAC/J efficiency \cite{miscuglio2018all}.

\section{Conclusion}

The optical activation functions made possible by the FPI have been shown to succeed in several pattern recognition applications, with the incoherent simulation of the network achieving a 98\% accuracy rate on the MNIST dataset, by using convolutional layers with higher stride in place of a dedicated pooling layer. Optimal performance has been shown to occur when the hardware constant of the FWHM, $\kappa$, is around 200 times the learning rate. The trainable activation parameter has been shown to simply require a initialisation variance of the same magnitude as $\kappa$ to allow maximum accuracies.

The complex valued coherent model has been tested and been shown be relatively insensitive to $\kappa$, requiring values of around 200 to 800 times the learning rate to ensure successful convergence. The activation function trainable parameter has been shown to be required to be initialised at 0, to avoid unstable training resulting in poor generalization. An MNIST accuracy of 95\% is achieved.

The maximum switching speed of current tunable FPIs currently imparts a high delay in the potential forward propagation speed of the network. However the rapidly growing field of MEMS based FPIs will result in higher switching speeds and therefore faster operation for the proposed ONN. New piezo-actuated FPIs are already showing higher switching speeds.

A free-space based ONN has been shown here to allow huge potential speed and power savings, and along with high scalability, could unlock many new applications where the required network or dataset size is prohibitively large for current digital system. As Moore's Law comes to an end, highly scalable hardware alternatives will be required to keep up with the rapid progression in deep learning research, and optics is a competitive contender that could allow this.

\section{Future Work}

Optical hardware solutions are a nascent field, and hardware specifications should be expected to dramatically improve over the design cycle time of new ONNs. The switching speed of the tunable FPI will decrease as soon as applications grow, as can be seen already with the piezo-actuated FPI \cite{antila2010mems}.

The use of multiple cascaded tunable FPIs at each neuron could allow a more customisable activation function at little loss of speed and compactness \cite{antila2010mems}. The architecture could be easily adapted to support multiple tunable parameters. With the extra globally optimizable parameter of the FWHM, some combination of FPIs could produce superior accuracies to those seen here.

GANs and RNNs are two examples of ANNs growing in popularity that require particularly large networks, with up to 85 million parameters \cite{Ledig_2017_CVPR, sak2014long}, that are well suited to ONNs.

Modern deep learning techniques could also be implemented in an ONN. Dropout could be implemented using DMDs in an ONN \cite{srivastava2014dropout}. Data augmentation could also be a technique particularly well suited to an ONN due to the high processing speed, where an enlarged dataset is not as prohibitive in terms of training time \cite{perez2017effectiveness, 6065487}.



\addcontentsline{toc}{section}{References}

\bibliography{sample}

\begin{thebibliography}{10}
\newcommand{\enquote}[1]{``#1''}

\bibitem{DeepLearning}
Y.~LeCun, Y.~Bengio, and G.~Hinton, \enquote{Deep learning,}
  {\protect\JournalTitle{Nature}} \textbf{521}, 436--44 (2015).

\bibitem{SilverHuangEtAl16nature}
D.~Silver, A.~Huang, C.~J. Maddison, A.~Guez, L.~Sifre, G.~van~den Driessche,
  J.~Schrittwieser, I.~Antonoglou, V.~Panneershelvam, M.~Lanctot, S.~Dieleman,
  D.~Grewe, J.~Nham, N.~Kalchbrenner, I.~Sutskever, T.~Lillicrap, M.~Leach,
  K.~Kavukcuoglu, T.~Graepel, and D.~Hassabis, \enquote{Mastering the game of
  {Go} with deep neural networks and tree search,}
  {\protect\JournalTitle{Nature}} \textbf{529}, 484--489 (2016).

\bibitem{mnih2015humanlevel}
V.~Mnih, K.~Kavukcuoglu, D.~Silver, A.~A. Rusu, J.~Veness, M.~G. Bellemare,
  A.~Graves, M.~Riedmiller, A.~K. Fidjeland, G.~Ostrovski, S.~Petersen,
  C.~Beattie, A.~Sadik, I.~Antonoglou, H.~King, D.~Kumaran, D.~Wierstra,
  S.~Legg, and D.~Hassabis, \enquote{Human-level control through deep
  reinforcement learning,} {\protect\JournalTitle{Nature}} \textbf{518},
  529--533 (2015).

\bibitem{Krizhevsky:2012:ICD:2999134.2999257}
A.~Krizhevsky, I.~Sutskever, and G.~E. Hinton, \enquote{Imagenet classification
  with deep convolutional neural networks,} in \emph{Proceedings of the 25th
  International Conference on Neural Information Processing Systems - Volume
  1,}  (Curran Associates Inc., USA, 2012), NIPS'12, pp. 1097--1105.

\bibitem{Esser11441}
S.~K. Esser, P.~A. Merolla, J.~V. Arthur, A.~S. Cassidy, R.~Appuswamy,
  A.~Andreopoulos, D.~J. Berg, J.~L. McKinstry, T.~Melano, D.~R. Barch,
  C.~di~Nolfo, P.~Datta, A.~Amir, B.~Taba, M.~D. Flickner, and D.~S. Modha,
  \enquote{Convolutional networks for fast, energy-efficient neuromorphic
  computing,} {\protect\JournalTitle{Proceedings of the National Academy of
  Sciences}} \textbf{113}, 11441--11446 (2016).

\bibitem{58356}
C.~{Mead}, \enquote{Neuromorphic electronic systems,}
  {\protect\JournalTitle{Proceedings of the IEEE}} \textbf{78}, 1629--1636
  (1990).

\bibitem{10.3389/fnins.2011.00108}
C.-S. Poon and K.~Zhou, \enquote{Neuromorphic silicon neurons and large-scale
  neural networks: Challenges and opportunities,}
  {\protect\JournalTitle{Frontiers in Neuroscience}} \textbf{5}, 108 (2011).

\bibitem{misra2010artificial}
J.~Misra and I.~Saha, \enquote{Artificial neural networks in hardware: A survey
  of two decades of progress,} {\protect\JournalTitle{Neurocomputing}}
  \textbf{74}, 239--255 (2010).

\bibitem{chen2017eyeriss}
Y.-H. Chen, T.~Krishna, J.~S. Emer, and V.~Sze, \enquote{Eyeriss: An
  energy-efficient reconfigurable accelerator for deep convolutional neural
  networks,} {\protect\JournalTitle{IEEE Journal of Solid-State Circuits}}
  \textbf{52}, 127--138 (2017).

\bibitem{graves2016hybrid}
A.~Graves, G.~Wayne, M.~Reynolds, T.~Harley, I.~Danihelka,
  A.~Grabska-Barwi{\'n}ska, S.~G. Colmenarejo, E.~Grefenstette, T.~Ramalho,
  J.~Agapiou \emph{et~al.}, \enquote{Hybrid computing using a neural network
  with dynamic external memory,} {\protect\JournalTitle{Nature}} \textbf{538},
  471 (2016).

\bibitem{AccelNNAnalogue}
N.~P. T. H. S. R. B. I. d. N. C. S. S. G. M. B. M. F. N. K. B. C. C. J.~Y.
  Ambrogio, S. and G.~Burr, \enquote{{Equivalent-accuracy accelerated
  neural-network training using analogue memory},}
  {\protect\JournalTitle{Nature}}  (2018).

\bibitem{miscuglio2018all}
M.~Miscuglio, A.~Mehrabian, Z.~Hu, S.~I. Azzam, J.~George, A.~V. Kildishev,
  M.~Pelton, and V.~J. Sorger, \enquote{All-optical nonlinear activation
  function for photonic neural networks,} {\protect\JournalTitle{Optical
  Materials Express}} \textbf{8}, 3851--3863 (2018).

\bibitem{vivien2012zero}
L.~Vivien, A.~Polzer, D.~Marris-Morini, J.~Osmond, J.~M. Hartmann, P.~Crozat,
  E.~Cassan, C.~Kopp, H.~Zimmermann, and J.~M. F{\'e}d{\'e}li,
  \enquote{Zero-bias 40gbit/s germanium waveguide photodetector on silicon,}
  {\protect\JournalTitle{Optics express}} \textbf{20}, 1096--1101 (2012).

\bibitem{yang2013multiplier}
R.~J. Lin~Yang, Lei~Zhang, \enquote{On-chip optical matrix-vector multiplier,}
  (2013).

\bibitem{40669}
H.~J. Caulfield, J.~Kinser, and S.~K. Rogers, \enquote{Optical neural
  networks,} {\protect\JournalTitle{Proceedings of the IEEE}} \textbf{77},
  1573--1583 (1989).

\bibitem{krishnamoorthy1992scalable}
A.~V. Krishnamoorthy, G.~Yayla, and S.~Esener, \enquote{A scalable
  optoelectronic neural system using free-space optical interconnects,}
  {\protect\JournalTitle{IEEE transactions on neural networks}} \textbf{3},
  404--413 (1992).

\bibitem{DBLP:journals/corr/abs-1901-03661}
S.~Colburn, Y.~Chu, E.~Shlizerman, and A.~Majumdar, \enquote{An optical
  frontend for a convolutional neural network,} {\protect\JournalTitle{CoRR}}
  \textbf{abs/1901.03661} (2019).

\bibitem{HybridCNNs}
J.~Chang, V.~Sitzmann, X.~Dun, W.~Heidrich, and G.~Wetzstein, \enquote{Hybrid
  optical-electronic convolutional neural networks with optimized diffractive
  optics for image classification,} {\protect\JournalTitle{Scientific Reports}}
  \textbf{8} (2018).

\bibitem{chen2016asp}
H.~G. Chen, S.~Jayasuriya, J.~Yang, J.~Stephen, S.~Sivaramakrishnan,
  A.~Veeraraghavan, and A.~Molnar, \enquote{Asp vision: Optically computing the
  first layer of convolutional neural networks using angle sensitive pixels,}
  in \emph{Proceedings of the IEEE Conference on Computer Vision and Pattern
  Recognition,}  (2016), pp. 903--912.

\bibitem{Hughes:18}
T.~W. Hughes, M.~Minkov, Y.~Shi, and S.~Fan, \enquote{Training of photonic
  neural networks through in situ backpropagation and gradient measurement,}
  {\protect\JournalTitle{Optica}} \textbf{5}, 864--871 (2018).

\bibitem{hamerly2018large}
R.~Hamerly, A.~Sludds, L.~Bernstein, M.~Solja{\v{c}}i{\'c}, and D.~Englund,
  \enquote{Large-scale optical neural networks based on photoelectric
  multiplication,} {\protect\JournalTitle{arXiv preprint arXiv:1812.07614}}
  (2018).

\bibitem{DeepLearningNanophotonicCircuits}
S.~S. M. P. T. B.-J. M. H. X. S. S. Z. H. L. D. E. . M.~S. Yichen~Shen,
  Nicholas C.~Harris, \enquote{Deep learning with coherent nanophotonic
  circuits,} {\protect\JournalTitle{Nature Photonics}} \textbf{11} (2017).

\bibitem{bueno2018reinforcement}
J.~Bueno, S.~Maktoobi, L.~Froehly, I.~Fischer, M.~Jacquot, L.~Larger, and
  D.~Brunner, \enquote{Reinforcement learning in a large-scale photonic
  recurrent neural network,} {\protect\JournalTitle{Optica}} \textbf{5},
  756--760 (2018).

\bibitem{FathomComputing}
\enquote{Fathom computing,}
  \url{https://web.archive.org/web/20190428165317/https://www.fathomcomputing.com/}.
  Accessed: 2019-04-28.

\bibitem{Lightelligence}
\enquote{Lightelligence - our technology,}
  \url{https://web.archive.org/web/20190201131606/https://www.lightelligence.ai/technology/}.
  Accessed: 2019-02-01.

\bibitem{Optalysys}
\enquote{Optalysys,}
  \url{https://web.archive.org/web/20190428165943/https://www.optalysys.com/}.
  Accessed: 2019-04-28.

\bibitem{2019arXiv190304579W}
I.~A.~D. {Williamson}, T.~W. {Hughes}, M.~{Minkov}, B.~{Bartlett}, S.~{Pai},
  and S.~{Fan}, \enquote{{Reprogrammable Electro-Optic Nonlinear Activation
  Functions for Optical Neural Networks},} {\protect\JournalTitle{arXiv
  e-prints}} arXiv:1903.04579 (2019).

\bibitem{wildfeuer2009resolution}
C.~F. Wildfeuer, A.~J. Pearlman, J.~Chen, J.~Fan, A.~Migdall, and J.~P.
  Dowling, \enquote{Resolution and sensitivity of a fabry-perot interferometer
  with a photon-number-resolving detector,} {\protect\JournalTitle{Physical
  Review A}} \textbf{80}, 043822 (2009).

\bibitem{Gorodetksy:10}
M.~L. Gorodetksy, A.~Schliesser, G.~Anetsberger, S.~Deleglise, and T.~J.
  Kippenberg, \enquote{Determination of the vacuum optomechanical coupling rate
  using frequency noise calibration,} {\protect\JournalTitle{Opt. Express}}
  \textbf{18}, 23236--23246 (2010).

\bibitem{hecht2013optics}
E.~Hecht, \emph{Optics: Pearson New International Edition} (Pearson Education
  Limited, 2013).

\bibitem{1057566}
L.~Cutrona, E.~Leith, C.~Palermo, and L.~Porcello, \enquote{Optical data
  processing and filtering systems,} {\protect\JournalTitle{IRE Transactions on
  Information Theory}} \textbf{6}, 386--400 (1960).

\bibitem{MAL-006}
Y.~Bengio, \enquote{Learning deep architectures for ai,}
  {\protect\JournalTitle{Foundations and Trends® in Machine Learning}}
  \textbf{2}, 1--127 (2009).

\bibitem{Long_2015_CVPR}
J.~Long, E.~Shelhamer, and T.~Darrell, \enquote{Fully convolutional networks
  for semantic segmentation,} in \emph{The IEEE Conference on Computer Vision
  and Pattern Recognition (CVPR),}  (2015).

\bibitem{goodfellow2014generative}
I.~Goodfellow, J.~Pouget-Abadie, M.~Mirza, B.~Xu, D.~Warde-Farley, S.~Ozair,
  A.~Courville, and Y.~Bengio, \enquote{Generative adversarial nets,} in
  \emph{Advances in neural information processing systems,}  (2014), pp.
  2672--2680.

\bibitem{2014arXiv1412.6806S}
J.~T. {Springenberg}, A.~{Dosovitskiy}, T.~{Brox}, and M.~{Riedmiller},
  \enquote{{Striving for Simplicity: The All Convolutional Net},}
  {\protect\JournalTitle{arXiv e-prints}} arXiv:1412.6806 (2014).

\bibitem{DBLP:journals/corr/abs-1710-05941}
P.~Ramachandran, B.~Zoph, and Q.~V. Le, \enquote{Searching for activation
  functions,} {\protect\JournalTitle{CoRR}} \textbf{abs/1710.05941} (2017).

\bibitem{sussillo2014random}
D.~Sussillo and L.~Abbott, \enquote{Random walk initialization for training
  very deep feedforward networks,} {\protect\JournalTitle{arXiv preprint
  arXiv:1412.6558}}  (2014).

\bibitem{glorot2010understanding}
X.~Glorot and Y.~Bengio, \enquote{Understanding the difficulty of training deep
  feedforward neural networks,} in \emph{Proceedings of the thirteenth
  international conference on artificial intelligence and statistics,}  (2010),
  pp. 249--256.

\bibitem{SemeionDataset}
\enquote{Semeion handwritten digit data set,}
  \url{https://web.archive.org/web/20190220185956/https://archive.ics.uci.edu/ml/datasets/semeion+handwritten+digit}.
  Accessed: 2019-02-20.

\bibitem{FourShapesDataset}
\enquote{16,000 images of four basic shapes (star, circle, square, triangle),}
  \url{https://web.archive.org/web/20190403184053/https://www.kaggle.com/smeschke/four-shapes}.
  Accessed: 2019-04-03.

\bibitem{ThreeShapesDataset}
\enquote{Babyaishapesdataset,}
  \url{http://www.iro.umontreal.ca/~lisa/twiki/bin/view.cgi/Public/BabyAIShapesDatasets}.
  Accessed: 2019-04-03.

\bibitem{Python}
\enquote{Python,}
  \url{https://web.archive.org/web/20190416213154/http://python.org/}.
  Accessed: 17/04/19.

\bibitem{NumPy}
\enquote{Numpy,}
  \url{https://web.archive.org/web/20190416233606/http://www.numpy.org/}.
  Accessed: 17/04/19.

\bibitem{TensorFlow}
\enquote{Tensorflow: An end-to-end open source machine learning platform,}
  \url{https://web.archive.org/web/20190416205518/https://www.tensorflow.org/}.
  Accessed: 17/04/19.

\bibitem{kingma2014adam}
D.~P. Kingma and J.~Ba, \enquote{Adam: A method for stochastic optimization,}
  {\protect\JournalTitle{arXiv preprint arXiv:1412.6980}}  (2014).

\bibitem{doi:10.1162/089976604773135104}
L.~Rosasco, E.~D. Vito, A.~Caponnetto, M.~Piana, and A.~Verri, \enquote{Are
  loss functions all the same?} {\protect\JournalTitle{Neural Computation}}
  \textbf{16}, 1063--1076 (2004).

\bibitem{IrisDataset}
\enquote{Iris data set,}
  \url{https://web.archive.org/web/20190213060058/https://archive.ics.uci.edu/ml/datasets/iris}.
  Accessed: 2019-02-15.

\bibitem{HeartDiseaseDataset}
\enquote{Heart disease uci dataset,}
  \url{https://web.archive.org/web/20190222145258/https://www.kaggle.com/ronitf/heart-disease-uci}.
  Accessed: 2019-02-22.

\bibitem{danihelka2016associative}
I.~Danihelka, G.~Wayne, B.~Uria, N.~Kalchbrenner, and A.~Graves,
  \enquote{Associative long short-term memory,} {\protect\JournalTitle{arXiv
  preprint arXiv:1602.03032}}  (2016).

\bibitem{941159}
T.~{Kim} and T.~{Adali}, \enquote{Complex backpropagation neural network using
  elementary transcendental activation functions,} in \emph{2001 IEEE
  International Conference on Acoustics, Speech, and Signal Processing.
  Proceedings (Cat. No.01CH37221),}  vol.~2 (2001), pp. 1281--1284 vol.2.

\bibitem{tygert2016mathematical}
M.~Tygert, J.~Bruna, S.~Chintala, Y.~LeCun, S.~Piantino, and A.~Szlam,
  \enquote{A mathematical motivation for complex-valued convolutional
  networks,} {\protect\JournalTitle{Neural computation}} \textbf{28}, 815--825
  (2016).

\bibitem{trabelsi2017deep}
C.~Trabelsi, O.~Bilaniuk, Y.~Zhang, D.~Serdyuk, S.~Subramanian, J.~F. Santos,
  S.~Mehri, N.~Rostamzadeh, Y.~Bengio, and C.~J. Pal, \enquote{Deep complex
  networks,} {\protect\JournalTitle{arXiv preprint arXiv:1705.09792}}  (2017).

\bibitem{arjovsky2016unitary}
M.~Arjovsky, A.~Shah, and Y.~Bengio, \enquote{Unitary evolution recurrent
  neural networks,} in \emph{International Conference on Machine Learning,}
  (2016), pp. 1120--1128.

\bibitem{jing2018gated}
L.~Jing, C.~Gulcehre, J.~Peurifoy, Y.~Shen, M.~Tegmark, M.~Soljacic, and
  Y.~Bengio, \enquote{Gated orthogonal recurrent units: On learning to forget,}
  in \emph{Workshops at the Thirty-Second AAAI Conference on Artificial
  Intelligence,}  (2018).

\bibitem{jose2017kronecker}
C.~Jose, M.~Cisse, and F.~Fleuret, \enquote{Kronecker recurrent units,}
  {\protect\JournalTitle{arXiv preprint arXiv:1705.10142}}  (2017).

\bibitem{mescheder2017numerics}
L.~Mescheder, S.~Nowozin, and A.~Geiger, \enquote{The numerics of gans,} in
  \emph{Advances in Neural Information Processing Systems,}  (2017), pp.
  1825--1835.

\bibitem{lecun1989generalization}
Y.~LeCun \emph{et~al.}, \enquote{Generalization and network design strategies,}
  in \emph{Connectionism in perspective,}  vol.~19 (Citeseer, 1989).

\bibitem{guberman2016complex}
N.~Guberman, \enquote{On complex valued convolutional neural networks,}
  {\protect\JournalTitle{arXiv preprint arXiv:1602.09046}}  (2016).

\bibitem{neuroptica}
\enquote{Neuroptica: Flexible simulation package for optical neural networks,}
  \url{https://web.archive.org/web/20190414221853/https://github.com/fancompute/neuroptica}.
  Accessed: 14/04/19.

\bibitem{CORREIA2000191}
J.~Correia, G.~de~Graaf, S.~Kong, M.~Bartek, and R.~Wolffenbuttel,
  \enquote{Single-chip cmos optical microspectrometer,}
  {\protect\JournalTitle{Sensors and Actuators A: Physical}} \textbf{82}, 191
  -- 197 (2000).

\bibitem{MicrospectrometerArray}
\enquote{Optical microspectrometer,}
  \url{https://web.archive.org/web/20190201124322/https://www.vttresearch.com/services/smart-industry/space-technologies/sensors-imaging-and-data-analysis/optical-microspectrometer}.
  Accessed: 2019-02-01.

\bibitem{MEMS-FPItunable}
\enquote{Ultra-compact near infrared spectrum sensor that integrates mems-fpi
  tunable filter and photosensor,}
  \url{https://web.archive.org/web/20190215143558/https://www.hamamatsu.com/resources/pdf/ssd/c13272-02_kacc1250e.pdf}.
  Accessed: 2019-02-15.

\bibitem{blomberg2010electrically}
M.~Blomberg, H.~Kattelus, and A.~Miranto, \enquote{Electrically tunable surface
  micromachined fabry--perot interferometer for visible light,}
  {\protect\JournalTitle{Sensors and Actuators A: Physical}} \textbf{162},
  184--188 (2010).

\bibitem{RISSANEN2012130}
A.~Rissanen, U.~Kantojärvi, M.~Blomberg, J.~Antila, and S.~Eränen,
  \enquote{Monolithically integrated microspectrometer-on-chip based on tunable
  visible light mems fpi,} {\protect\JournalTitle{Sensors and Actuators A:
  Physical}} \textbf{182}, 130 -- 135 (2012).

\bibitem{antila2010mems}
J.~Antila, A.~Miranto, J.~M{\"a}kynen, M.~Laamanen, A.~Rissanen, M.~Blomberg,
  H.~Saari, and J.~Malinen, \enquote{Mems and piezo actuator-based fabry-perot
  interferometer technologies and applications at vtt,} in
  \emph{Next-Generation Spectroscopic Technologies III,}  vol. 7680
  (International Society for Optics and Photonics, 2010), p. 76800U.

\bibitem{arbabi2015dielectric}
A.~Arbabi, Y.~Horie, M.~Bagheri, and A.~Faraon, \enquote{Dielectric
  metasurfaces for complete control of phase and polarization with
  subwavelength spatial resolution and high transmission,}
  {\protect\JournalTitle{Nature nanotechnology}} \textbf{10}, 937 (2015).

\bibitem{MicromirrorArray}
\enquote{Texas instruments digital micromirror devices,}
  \url{https://web.archive.org/web/20190201123745/http://www.ti.com/lit/an/dlpa008b/dlpa008b.pdf}.
  Accessed: 2019-02-01.

\bibitem{Ledig_2017_CVPR}
C.~Ledig, L.~Theis, F.~Huszar, J.~Caballero, A.~Cunningham, A.~Acosta,
  A.~Aitken, A.~Tejani, J.~Totz, Z.~Wang, and W.~Shi, \enquote{Photo-realistic
  single image super-resolution using a generative adversarial network,} in
  \emph{The IEEE Conference on Computer Vision and Pattern Recognition (CVPR),}
   (2017).

\bibitem{sak2014long}
H.~Sak, A.~Senior, and F.~Beaufays, \enquote{Long short-term memory recurrent
  neural network architectures for large scale acoustic modeling,} in
  \emph{Fifteenth annual conference of the international speech communication
  association,}  (2014).

\bibitem{srivastava2014dropout}
N.~Srivastava, G.~Hinton, A.~Krizhevsky, I.~Sutskever, and R.~Salakhutdinov,
  \enquote{Dropout: a simple way to prevent neural networks from overfitting,}
  {\protect\JournalTitle{The Journal of Machine Learning Research}}
  \textbf{15}, 1929--1958 (2014).

\bibitem{perez2017effectiveness}
L.~Perez and J.~Wang, \enquote{The effectiveness of data augmentation in image
  classification using deep learning,} {\protect\JournalTitle{arXiv preprint
  arXiv:1712.04621}}  (2017).

\bibitem{6065487}
D.~C. Ciresan, U.~Meier, L.~M. Gambardella, and J.~Schmidhuber,
  \enquote{Convolutional neural network committees for handwritten character
  classification,} in \emph{2011 International Conference on Document Analysis
  and Recognition,}  (2011), pp. 1135--1139.

\end{thebibliography}






\end{document}